\documentclass{aastex}
\usepackage{emulateapj5}

\newcommand{\kms}{km\thinspace s$^{-1}$}
\newcommand{\ergss}{ergs\thinspace s$^{-1}$}
\newcommand{\ergsscm}{ergs\thinspace s$^{-1}$\thinspace cm$^{-2}$}
\def\Msun{\hbox{$\thinspace M_{\odot}$}}
\def\Zsun{\hbox{$\thinspace Z_{\odot}$}}

\slugcomment{accepted for publication in The Astrophysical Journal Letters}

\shortauthors{Zepf et al.}

\begin{document}


\title{Very Broad [\ion{O}{3}]$\lambda\lambda$4959, 5007
Emission from the NGC~4472 
Globular Cluster RZ2109 and Implications for the Mass of Its Black Hole 
X-ray Source\altaffilmark{1}}


\author{Stephen E. Zepf\altaffilmark{2}, 
Daniel Stern\altaffilmark{3},
Thomas J. Maccarone\altaffilmark{4}, 
Arunav Kundu\altaffilmark{2},
Marc Kamionkowski\altaffilmark{5},
Katherine L. Rhode\altaffilmark{6}, 
John J. Salzer\altaffilmark{6,7}, 
Robin Ciardullo\altaffilmark{8},
Caryl Gronwall\altaffilmark{8}}

\altaffiltext{1} {Based on observations made at the W. M. Keck Observatory, 
which is operated as a scientific partnership among the California 
Institute of Technology, the University of California, and the 
National Aeronautics and Space Administration. The Observatory was 
made possible by the generous financial support of the W. M. Keck Foundation.}
\altaffiltext{2}{Department of Physics \& Astronomy, Michigan State University,
East Lansing, MI 48824; e-mail: zepf, akundu@pa.msu.edu}
\altaffiltext{3}{Jet Propulsion Laboratory,
California Institute of Technology, Pasadena, CA 91109; 
e-mail: stern@zwolfkinder.jpl.nasa.gov}
\altaffiltext{4}{School of Physics \& Astronomy, University of Southampton, 
Southampton, Hampshire SO17 1BJ; e-mail: tjm@astro.soton.ac.uk}
\altaffiltext{5}{California Institute of Technology, Mail Code 130-33,
Pasadena, CA 91125; e-mail: kamion@tapir.caltech.edu}
\altaffiltext{6}{Department of Astronomy, Indiana University,
Bloomington, IN 47405; e-mail: rhode, slaz@astro.indiana.edu}
\altaffiltext{7}{Department of Astronomy, Wesleyan University,
Middletown, CT 06459}
\altaffiltext{8}{Department of Astronomy and Astrophysics, Penn State 
University, University Park, PA 16802; e-mail:caryl, rbc@astro.psu.edu}

\begin {abstract}

We present Keck LRIS spectroscopy of the black hole-hosting 
globular cluster RZ2109 in the Virgo elliptical galaxy NGC~4472. 
We find that this object has extraordinarily broad 
[\ion{O}{3}]$\lambda$5007 and [\ion{O}{3}]$\lambda$4959 emission 
lines, with velocity widths of approximately 2,000\kms . 
This result has significant implications for the nature 
of this accreting black-hole system and the mass of the  
globular cluster black hole. We show that the broad 
[\ion{O}{3}]$\lambda$5007 emission must arise from material 
driven at high velocity from the black hole system.
This is because the volume available near the black hole  
is too small by many orders of magnitude to have enough [OIII]  
emitting atoms to account for the observed 
L([\ion{O}{3}]$\lambda$5007) at high velocities, even if this 
volume is filled with Oxygen 
at the critical density for [\ion{O}{3}]$\lambda$5007.
The Balmer emission is also weak, indicating the 
observed [OIII] is not due to shocks. We therefore conclude
that the [\ion{O}{3}]$\lambda\lambda$4959,5007 
is produced by photoionization of material driven across the cluster.
The only known way to drive significant material 
at high velocity is for a system accreting mass near or above 
its Eddington limit, which indicates a stellar mass black hole. 
Since it is dynamically implausible to form an accreting stellar mass
black hole system in a globular cluster with an intermediate
mass black hole (IMBH), it appears this massive globular 
cluster does {\it not} have an IMBH. We discuss further tests 
of this conclusion, and its implications for the 
$M_{BH} - M_{\rm stellar}$ and $M_{BH} - \sigma$ relations.

\end{abstract}


\section{Introduction}

Maccarone et al. (2007) presented the first unambiguous
evidence for a black hole in a globular cluster.
This was based on an XMM observation of strong variability 
in a high luminosity X-ray source ($\simeq 4 \times 10^{39}$ ergs/s)
located in the spectroscopically confirmed globular
cluster RZ2109 in the Virgo elliptical galaxy NGC~4472.
The high X-ray luminosity, about an order of
magnitude greater than the Eddington luminosity
of a neutron star, requires either a black hole
or multiple neutron stars in the old stellar population
of the globular cluster. The strong variability
rules out multiple neutron stars as the source
of the bright X-ray emission, and thus indicates
the globular cluster RZ2019 hosts a black hole X-ray
source. 

Subsequently, Zepf et al.\ (2007, hereafter Z2007) discovered
broad [\ion{O}{3}]$\lambda$5007 emission from this black hole-hosting 
globular cluster. Specifically, Z2007 analyzed existing 
VLT spectra of this object covering a wavelength range of
$5000 - 5800$ \AA . They identified a broad
emission line in these data at a wavelength in very close 
agreement (within 10 \kms) with that expected for 
[\ion{O}{3}]$\lambda$5007 at the radial velocity of the globular cluster 
previously determined from its stellar 
absorption lines. Z2007 also found that the
[\ion{O}{3}]$\lambda$5007 emission line was much broader than any
possible velocity dispersion for the globular
cluster, with an estimated width of several
hundred \kms, and the possibility of much broader
wings in the data. They noted such a large velocity 
width clearly points to the black hole as the source 
of the power driving the nebular emission.

As described in Z2007, understanding
how the broad [\ion{O}{3}]$\lambda$5007 emission is produced
provides a way to constrain the mass of the black
hole in this globular cluster. One possibility
is that the black hole is a stellar mass system,
accreting near or above its Eddington limit.
Such systems can drive strong, high velocity
winds (Begelman, King, \& Pringle 2006 and references therein).
In this case, broad [OIII] could be produced
either by shocks as the wind passes through the
globular cluster's interstellar medium,
or by photoionization of this wind material
across the cluster. Alternatively, it could
be an intermediate mass black hole, in which
case it would be far below its Eddington
luminosity given the observed X-ray luminosity.
In this case, no strong winds would be driven,
and any broad [OIII] emission would have to be
produced by photoionizing of material
in very close proximity to the black hole.

The X-ray data alone do not clearly distinguish
between these possibilities. There are also
a wide range of theoretical predictions for
the fate and properties of black holes in
globular clusters. For example, extrapolations
of the $M_{BH} - \sigma$ and the 
$M_{BH} - M_{\rm stellar}$ relations tend to
predict IMBHs in globular clusters (see Section 4),
and some models find sufficiently dense young
globular clusters cores may experience runaway
stellar mergers that may form IMBHs (e.g.\
Portegies Zwart \& McMillan 2002,
Freitag et al.\ 2006, 
but see also Yungelson et al.\ 2008). 
On the other hand, as stellar mass 
black holes sink to the centers
of globular clusters through mass segregation, 
dynamical interactions between them may eject most
or all of the black holes from the core, either
into the outskirts of the globular cluster 
(Mackey et al.\ 2007), or out of the cluster completely 
(e.g.\ Sigurddson \& Hernquist 1993, Kulkarni et al.\ 1993). 

In this {\em Letter}, we present new optical spectroscopy
aimed at understanding the physical nature of the accreting
black hole source in the globular cluster RZ2109.
The optical spectra used in Z2007
were available because of an ongoing kinematic
study of the NGC~4472 globular cluster system using FLAMES,
a multi-fiber spectrograph on the VLT.
The FLAMES spectra have very limited wavelength coverage
and a modest signal-to-noise ratio.
The work presented here both 
significantly increases the wavelength coverage to 
investigate diagnostic line ratios, and improves the
signal to noise ratio to better constrain the width
of the broad [\ion{O}{3}]$\lambda$5007 emission line.

\section{Observations}

We obtained optical spectra of the NGC~4472 globular cluster RZ2019, which
contains the X-ray source XMMU 1229397+07533, on UT 2007 December 17$-$18
with the Low Resolution Imaging Spectrometer (LRIS; Oke et al.\ 1995)
on the Keck I telescope.  The observations, taken at a position angle
of $74.4\deg$, through a 1\farcs5 slit, used the D560 dichroic, the 400
lines mm$^{-1}$ grism ($\lambda_{\rm blaze} = 3400 {\rm \AA}$) on the
blue side, and the 400 lines mm$^{-1}$ grating ($\lambda_{\rm blaze} =
8500 {\rm \AA}$) on the red side.  The spectral resolution for objects
filling the slit was $R \sim 400$ and 700, on the blue and 
red sides respectively.  The nights
were not photometric, and the seeing ranged from moderate (1\farcs2;
first night) to poor (1\farcs6; second night).  The exposure times were
1500~s each night.

Data processing followed standard slit spectroscopy procedures.
In particular, we extracted the spectra using a 1\farcs5 wide aperture and
calculated the spectral dispersions using lamp spectra obtained earlier
in the nights, adjusting the wavelength zero point based on telluric
emission lines.  
The spectra were flux calibrated using observations
of standard stars from Massey \& Gronwall (1990) obtained during the
same nights.  Since the nights were not photometric, the spectra were
scaled to match the observed optical brightness of RZ2109 ($V = 21.0$).  
We also corrected the atmospheric A and B band bands based on the 
standard star spectra. The final, calibrated spectrum is shown in Figure 1.

\section {Results}
In this letter, we focus on two key aspects of the 
remarkable spectrum of RZ2109 shown in Figure 1. 
These are - 1) the
very broad velocity width of the [\ion{O}{3}]$\lambda\lambda$4959, 5007,
and 2) the [\ion{O}{3}]$\lambda$5007 line luminosity and the
large ratio of the [\ion{O}{3}]$\lambda$5007 to H$\beta$. 
We also note that this spectrum clearly shows
[\ion{O}{3}]$\lambda$5007, [\ion{O}{3}]$\lambda$4959, 
and likely [\ion{O}{3}]$\lambda$4363 in
emission at the radial velocity of the globular cluster 
previously determined
from its stellar absorption lines. This
confirms beyond any doubt that these emission
lines are from the globular cluster RZ2109,
as found by Z2007 from the [\ion{O}{3}]$\lambda$5007
line alone.

\subsection {The Extraordinary [\ion{O}{3}]$\lambda\lambda$4959, 5007 
Velocity Width}
The spectrum in Figure 1 shows that the
[\ion{O}{3}]$\lambda$4959 and $\lambda$5007 emission lines are
extremely broad, so broad that the lines
are blended. To determine the velocity
width of these lines we first use standard deblending
procedures. Specifically, we fit the two lines with
two Gaussians, keeping the width of the lines
the same because they come from the same initial atomic 
state. This gives a FWHM of the lines 
of 33 \AA , corresponding to a velocity FWHM of 
approximately 2000 \kms . 

Figure 1 shows the lines have stronger central
peaks and broader wings than the best fitting
Gaussian functions. We therefore also calculate the
velocity width of the lines by determining the longest 
wavelength at which the [\ion{O}{3}]$\lambda$5007 line clearly
has significant flux. We determine the corresponding
shortest wavelength for the [\ion{O}{3}]$\lambda$4959 line 
to estimate the lowest velocity component.
Using this technique, we find a velocity extent
of the emission of at least $+1600$ \kms\ and $-1400$ \kms .
Whether the [\ion{O}{3}]$\lambda\lambda$4959, 5007 lines are described as 
having FWHM of 33 \AA , or having a velocity extent 
of at least $\pm 1500$ \kms,
these lines are extraordinarily broad, as broad
as almost any [\ion{O}{3}]$\lambda$5007 lines observed,
while being located in a globular cluster
with an estimated mass of a few million
solar masses and an escape velocity
several orders of magnitude less than
the observed velocity width.

\subsection{Line Luminosities and Ratios}

The [\ion{O}{3}]$\lambda$5007 line luminosity and its comparison
to other diagnostic lines such as H$\beta$ place
important constraints on the nature of the accreting
black hole system.
One way to calculate L([\ion{O}{3}]$\lambda$5007) is
to use the Gaussian fits to the [OIII]
lines described above. These give fluxes of 
$1.8 \times 10^{-16}$ \ergss cm$^{-2}$ and 
$4.4 \times 10^{-16}$ \ergss cm$^{-2}$
for [OIII]4949 and [\ion{O}{3}]$\lambda$5007 respectively. 
Adopting a 16 Mpc distance for NGC~4472, then
gives a line luminosity of L([\ion{O}{3}]$\lambda$5007)$ = 
1.4 \times 10^{37}$ \ergss .

An alternative approach to determining the 
[\ion{O}{3}]$\lambda$5007 luminosity that avoids Gaussian fitting 
is to determine the total [OIII]$\lambda 4959 + \lambda 5007$
luminosity, and adopt the 3 to 1 ratio of L([\ion{O}{3}]$\lambda$5007) 
to L([\ion{O}{3}]$\lambda$4959) given by atomic physics 
(e.g.\ Dimitrijevi\'c et al.\ 2006, Storey \& Zeippen 2000).
We do this by measuring the straightforward equivalent
width and total [OIII]$\lambda 4959 + \lambda 5007$ flux, 
and then using the above ratio between
the lines to determine L([\ion{O}{3}]$\lambda$5007).
This calculation gives an [\ion{O}{3}]$\lambda$5007 equivalent width
of 31 \AA , and L([\ion{O}{3}]$\lambda$5007) $= 1.4 \times 10^{37}$ \ergss .
While the close agreement of this determination with that
above is undoubtedly partially fortuitous, it is also reassuring
that different techniques for determining the line
luminosities give similar answers.
We also note that the L([\ion{O}{3}]$\lambda$5007) found here
is significantly higher than the best estimate in Z2007.
Z2007 noted that the line might have
broad wings, and that if the broad wings were real, 
their velocity width and L([\ion{O}{3}]$\lambda$5007) estimates would
be lower limits.

The [\ion{O}{3}]$\lambda$5007 to H$\beta$ ratio is a valuable diagnostic 
emission line mechanism. It is immediately
apparent from Figure 1 that the Balmer line emission
is not strong; the Balmer lines are all seen in
absorption. Balmer absorption lines are expected
from the stellar population of the globular cluster.
The H$\beta$ emission can be estimated by comparing 
the amount of H$\beta$ absorption
expected from stellar population models to that
observed, attributing any difference to filling in
of the stellar absorption line by H$\beta$ emission
from the black hole system. To carry out this
measurement, we use the stellar populations
models of Bruzual \& Charlot (2003), and those of
Vazdekis et al.\ (2008) based on the spectral library 
of S\'anchez-Bl\'azquez et al. (2006), since these 
provide spectra at fairly high resolution.
For each set of models, we adopt an age of 12 Gyr
and set the model metallicity to match the observed 
optical colors of RZ2019. We then 
analyze the spectral output of the models in the same 
way as the data. Both models predict an equivalent width
of 2.6 \AA\ for the H$\beta$ absorption line,
while the same measurement of the observed spectrum
gives 1.5 \AA . 

Our analysis therefore suggests an
H$\beta$ equivalent width of about 1.1 \AA\,
giving a very rough L[\ion{O}{3}]$\lambda$5007/L[H$\beta$] ratio of 30.
We caution that this is somewhat dependent on the
stellar populations modeling of the stellar Balmer
absorption. There are also hints in the
current data that the Balmer emission may not be
as broad as the [\ion{O}{3}]$\lambda\lambda$4959, 5007 lines, and
we plan to address this and other questions about the detailed
shape of the [\ion{O}{3}]$\lambda\lambda$4959, 5007 line complex in
future work with higher spectral resolution.
we will address this further in a future paper.
Here we simply note that the [\ion{O}{3}]$\lambda$5007/H$\beta$ ratio
appears to be large.

\section {Discussion}
Zepf et al. (2007) noted three possibilities for
producing the broad [\ion{O}{3}]$\lambda$5007 emission.
One is that it could be produced by shocks from
a strong, high velocity wind driven from a
black hole. Driving such winds requires a
system undergoing mass transfer at or above
its Eddington limit, 
which given the observed $L_{X} \simeq 4 \times 10^{39}$
\ergss , would indicate a stellar mass black hole system
of $M_{BH} \simeq 10$\Msun.
Alternatively, the velocity width could be generated
by a strong wind driven across the globular cluster,
but the ionization source could be photoionization
from the powerful central system. Because this case
involves strong winds, it too implicates a stellar
mass black hole system. The third possibility
is that the high velocities are due to gravitational
motions very close to the central black hole, and the
material is photoionized. This scenario likely
implicates an IMBH to more readily give the high velocities
at sufficient distances to allow [OIII] emission.

The new results presented here -- the very broad
[\ion{O}{3}]$\lambda$4959 and $\lambda$5007 emission 
lines extending over more than one thousand \kms, and 
the large [\ion{O}{3}]$\lambda$5007/H$\beta$ ratio --
strongly constrain these possibilities.
First, the large [OIII]/H$\beta$ ratio
indicates the [\ion{O}{3}]$\lambda$5007 
is not produced directly by shocks.
Secondly, the large line widths
extending to high velocities rule out the idea
the velocity widths seen in the 
[\ion{O}{3}]$\lambda\lambda$4959, 5007
lines can be due to gravitational motions
very near a black hole. This is because the volume
available near the black hole where the line widths
would be as large as observed is too small by orders
of magnitude to have enough [OIII] emitting atoms.

To calculate the maximum possible [\ion{O}{3}]$\lambda$5007 flux 
for material orbiting around a black hole, we first find the volume 
available in a shell corresponding to a given circular velocity 
around the black hole. We then determine the maximum possible 
[\ion{O}{3}]$\lambda$5007 flux at this velocity by assuming 
this volume is filled completely with material at the critical 
density of [\ion{O}{3}]$\lambda$5007 with a solar O/H ratio, and 
taking the collisional excitation rate 
for [\ion{O}{3}]$\lambda$5007 
(Osterbrock \& Ferland 2006). 
We find that the maximum flux for material with a velocity 
of 1500 \kms around a 1000\Msun\ black hole is 
$F_{\lambda}$([\ion{O}{3}]$\lambda$5007) $= 8 \times 10^{-26}$ \ergsscm\AA$^{-1}$.
This is the maximal flux in that it assumes the entire
available volume is at the critical density of [\ion{O}{3}]$\lambda$5007,
that all the O atoms are doubly ionized, and that the O/H
ratio is solar, while the best estimate for the metallicity
of RZ2109 is about 1/50 \Zsun\ (Maccarone et al.\ 2007, Rhode \& Zepf 2001).
In contrast to this maximum flux from material orbiting around
a black hole, the observed $F_{\lambda}$([\ion{O}{3}]$\lambda$5007) at
velocities around 1500 \kms\
is somewhat larger than $10^{-18}$ \ergsscm\AA$^{-1}$ (see Figure 1),
or more than $10^7$ times greater.
Thus, even the maximal emission from material orbiting
around a black hole fails by many orders of magnitude to
match the flux in the observed broad [\ion{O}{3}]$\lambda$5007 line.
One can increase the flux by assuming an 
even higher O abundance, or by increasing
the available volume by going to higher $M_{BH}$ or lower velocity.
However, the difference is so great and the presence of significant
flux at large velocities clearly observed, that these do not
begin to reconcile this scenario with the observations.

Therefore, the only scenario consistent with the data is
that in which the black hole system drives material
across the cluster, and that this material is also
photoionized by the central X-ray source.
Such a scenario strongly implies that the black hole
is a stellar mass black hole and not an IMBH,
based on both theoretical and observational considerations.
Extensive theoretical work has found that strong winds
like those found here are driven only by systems with
mass accretion rates similar to or larger than the 
Eddington limit (e.g.\ Proga 2007 and references therein).
Given the observed X-ray luminosity of $4 \times 10^{39}$ \ergss ,
this strongly implies a stellar mass black hole, with a
mass of $M_{BH} \simeq 10$\Msun\ giving a good fit to the 
X-ray data in this scenario (Maccarone et al.\ 2007).

A stellar mass for the accreting black hole system
in RZ2109 is also supported by the high $L/L_{\rm Edd}$ found
for the small fraction of extragalactic AGN with observed 
broad velocity widths in [\ion{O}{3}]$\lambda$5007.
One well-studied example is NGC~1068, which has long been known
to have highly blueshifted [\ion{O}{3}]$\lambda$5007 components 
with the emission dominated 
by photoionization (Groves et al.\ 2004, Kinkhabwala et al.\ 2002), 
and for which the central source
appears to be above its formal Eddington limit 
(e.g.\ Bock et al.\ 2000). A small fraction of 
narrow-line Seyfert 1 galaxies are also observed to have 
either an underlying broad or significantly blueshifted 
[\ion{O}{3}]$\lambda$5007
component, which is found to be related to high $L/L_{\rm Edd}$ 
ratios (e.g.\ Komossa et al.\ 2008, Boroson 2005),
further suggesting RZ2109 is near its Eddington luminosity. 

Thus, theory and observation indicate that the
black hole in this globular cluster
system is very likely a stellar mass black hole.
As discussed in Z2007, it seems highly unlikely
that a close, accreting black hole binary system
could form in a globular cluster with an IMBH at its center. 
This is because the dynamical processes by which close
black hole X-ray binaries form in globular clusters
happen in their cores, but dynamically, 
a globular cluster core containing an IMBH will have typically 
ejected its stellar mass black holes (see Z2007 for
full discussion).

The globular cluster RZ2109 appears to have only
stellar mass black hole(s), with no intermediate 
mass black hole. As such it falls far off the 
relations between black hole mass and the 
mass and velocity dispersion of the stellar 
system in which it is found. To show this
explicitly, we first calculate the $M_{BH}$
for RZ2109 implied by the $M_{BH} - M_{\rm stellar}$ 
relation found by H\"aring \& Rix (2004). For the 
stellar mass of RZ2109, we use its absolute luminosity 
of $M_V = -10$, and a $(M/L)_V =2$ standard
for globular clusters to estimate a stellar
mass of about $2 \times 10^6$\Msun. Based on the
H\"aring \& Rix (2004) relation, a $M_{BH}$ of
900\Msun\ would be predicted for RZ2109, with an 
uncertainty of a factor of two. This is far greater 
than the stellar mass of $\sim 10$\Msun\ indicated 
by the calculations above.
Extrapolating the $M_{BH} - \sigma$ relation to
RZ2109 tends to predict even larger black hole
masses than the $M_{BH} - M_{\rm stellar}$ relation. 
A direct determination of the $\sigma$ of RZ2109 is 
not feasible, but we can compare it to the $\sigma$
found for Galactic globular clusters of similar
absolute magnitude. These are the luminous
clusters such as 47 Tuc, Omega Cen,
and M54, which have $\sigma$ ranging from
about $12-20$ \kms. Adopting a typical
value of 16 \kms, the Tremaine et al.\ (2002)
$M_{BH} - \sigma$ relation gives a predicted
central black hole mass in RZ2109 of 5000\Msun .

The present analysis would seem to indicate that not all
old, metal-poor stellar systems form black holes
consistent with $M_{BH} - M_{\rm stellar}$ and 
$M_{BH} - \sigma$ relations found for more massive
galaxies. 
It also indicates 
that stellar dynamics did not lead to coalesence 
into an intermediate mass black hole, at least in 
this one fairly massive globular cluster.
More speculatively, the results suggest that the dark 
matter halos
that presumably surround the early galaxies which grow
hierarchically to make massive galaxies at the current
epoch are important for retaining
gas which is subsequently accreted to fuel the BH growth
and set the $M_{BH} - \sigma$ relation.
To further test these conclusions, additional 
work on forming accreting stellar mass black hole systems in globular 
clusters with and without an IMBH would be valuable, as would further 
understanding of winds driven from systems undergoing mass transfer 
near or above their Eddington limit. Data with higher
spectral resolution would allow a more detailed study of the
velocity structure in the [\ion{O}{3}]$\lambda\lambda$4959, 5007
emission lines which would help constrain the geometry of the
emitting material.
We also note that the extended nature of the [\ion{O}{3}]$\lambda$5007
emission in RZ2109 can be directly tested 
by spatially resolved spectroscopy with HST.

\acknowledgments
SEZ acknowledges support for this work from
NSF award AST-0406891, AK acknowledges support from
NASA-LTSA grant NAG5-12975, the work of DS was carried out 
at Jet Propulsion Laboratory, California Institute
of Technology, under a contract with NASA.
We also thank the referee for a detailed and careful report.
We wish to acknowledge the significant cultural role that 
the summit of Mauna Kea plays within the indigenous Hawaiian community; 
we are fortunate to have the opportunity to conduct observations 
from this mountain.

\clearpage

\begin{figure}
\centerline{\includegraphics[angle=-90,width=\textwidth]{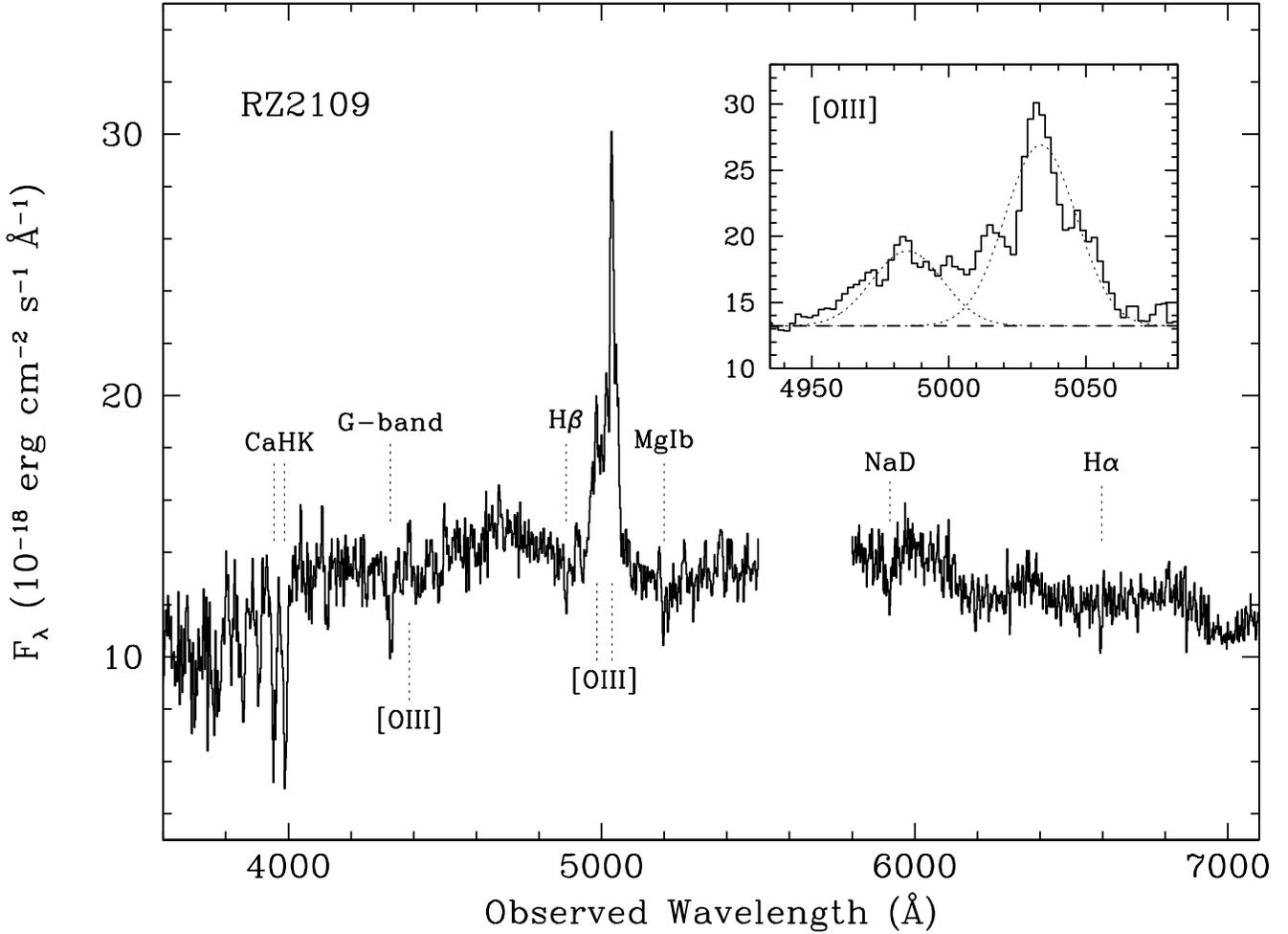}}
\caption{
A plot of $F_{\lambda}$ vs. $\lambda$ for RZ2109, the NGC~4472 
globular cluster in which the black hole system XMMU 1229397+075333 
is found. This figure shows the remarkably strong and broad 
[\ion{O}{3}]$\lambda\lambda$4959, 5007 [OIII] emission lines. 
The upper inset figure shows the [\ion{O}{3}]$\lambda\lambda$4959, 5007
region in more detail. The light dashed lines on the inset figure
are the Gaussian fits to the [\ion{O}{3}]$\lambda\lambda$4959, 5007 lines, 
with a FWHM of 2,000 \kms . This demonstrates the extraordinary width
of these lines. The paper focuses on these lines and the implication
of their luminosity and velocity width. There also appears to be
velocity structure beyond the individual Gaussian fits, but higher
spectral resolution than the $R \simeq 400$ of these data
is needed for a reliable assessment of structure within the very
broad lines. The gap around 5600 \AA\ is due
to the dichroic in the LRIS spectrograph. The stellar absorption
lines are from the stellar population of the
host globular cluster, with some likely filling in of the Balmer lines
due to weak Balmer emission, as discussed in the text.
\label{fig1}
}

\end{figure}

\end{document}